# A Stochastic Planning Method for Low-carbon Building-level Integrated Energy System Considering Electric-Heat-V2G Coupling☆

Zuxun Xiong [a], Xinwei Shen [a,b,*], Qinglai Guo [a,c], Hongbin Sun [a,c,*]

[a] Tsinghua-Berkeley Shenzhen Institute, Tsinghua Shenzhen International Graduate School, Tsinghua University, Shenzhen 518055, China
[b] Institute for Ocean Engineering, Tsinghua Shenzhen International Graduate School, Tsinghua University, Shenzhen 518055, China
[c] Department of Electrical Engineering, Tsinghua University, Beijing 100084, China

*Abstract--* **The concept of low-carbon building is proposed to ameliorate the climate change caused by environmental problems and realize carbon neutrality at the building level in urban areas. In addition, renewable energy curtailment in the power distribution system, as well as low efficiency due to independent operation of traditional energy systems, has been addressed by the application of integrated energy system (IES) to some extent. In this paper, we propose a planning method for low-carbon building-level IES, in which electric vehicles (EV) and the mode of Vehicle to Grid (V2G) are considered and further increase the flexibility of low-carbon buildings. The proposed planning model optimize the investment, operation costs and CO2 emission for building-level IES, so as to achieve the maximum benefit of the construction of the low-carbon building and help the realization of carbon neutrality. Moreover, we consider the uncertainty of distributed renewable energy, multi-energy load fluctuation and the random behavior of EV users, then formulating a two-stage stochastic programming model with chance constraints, in which heuristic moment matching scenario generation (HMMSG) and sample average approximation (SAA) method are applied. In case study, a real IES commercial building in Shanghai, where photovoltaic (PV), energy storage system (ESS), fuel cell (FC), EV, etc. are included as planning options, is used as numerical example to verify the effectiveness of the proposed planning method, with functions of ESS and EV in IES are analyzed in detail in different operation scenarios.**

## Nomenclature

*Abbreviations:*

| | |
|---|---|
| IES | Integrated Energy System |
| EV | Electric Vehicle |
| V2G | Vehicle to Grid |
| HMMSG | Heuristic Moment Matching Scenario Generation |
| SAA | Sample Average Approximation |
| ESS | Energy Storage System |
| FC | Fuel Cell |
| CCTS | Chance-Constrained Two-Stage |
| EH | Energy Hub |
| SOFC | Solid Oxide Fuel Cell |
| PEM | Polymer Electrolyte Membrane |
| PV | Photovoltaic |
| SOC | State of Charge |
| P2G | Power to gas |

*Sets:*

| | |
|---|---|
| $\Omega$ | Set of fuel cell to be selected |
| $\Psi_{parking}$ | Set of the parking period of EV |

*Indices:*

| | |
|---|---|
| $i$ | Index of different kind of FC |
| $t$ | Index of time / hour |
| $j$ | Index of EVs |
| $e,h,g,l$ | Index of different kind of energy |
| $s$ | Index of scenario |
| $ch/dis$ | Index of charge/discharge |

*Parameters:*

| | |
|---|---|
| $IC$ | Investment cost of equipment |
| $m$ | Coefficient to convert costs |
| $\gamma$ | Discount rate of cash |
| $PP$ | Planning period |
| $N$ | Number of scenarios |
| $r$ | Cost of electricity/gas |
| $C^{tax}$ | Tax for carbon emission |
| $\varepsilon$ | Emission coefficient of fuel/grid |
| $p$ | Penalty of substandard SOC level |
| $NEV$ | Number of EVs |
| $\mathbf{L}$ | Load of the building-level IES |
| $\mathbf{C}$ | Coupling matrix of EH model |
| $\eta$ | Charge/Discharge efficiency of ESS/EV |
| $\alpha$ | Charge/Discharge efficiency of ESS/EV |
| $T_{ch}^{BESS}$ | Maximum full charge times of BESS |
| $\varsigma$ | Proportion of substandard scenarios |
| $P_{max}^{Grid}$ | Maximum power of the grid |
| $P_{i,max}^{FC}$ | Maximum power of $i^{th}$ FC |
| $P_{charger}^{EV}$ | Power of EV charger |
| $P_{max}^{PV}$ | Maximum power of the PV |
| $E_{max}^{EV}$ | Capacity of EV |

*Variables:*

| | |
|---|---|
| $\mathbf{X,P,Y,Z}$ | Vector/Matrix of decision variables |
| $X^{ESS}$ | Capacity of ESS that will be invested, unit: kWh |
| $X_i^{FC}$ | Numbers of $i^{th}$ FC that will be invested, unit: sets |
| $P_{s,t}^{Grid}$ | Output of the grid at time $t$ in scenario $s$, unit: kW |

☆ The short version of this paper was presented at virtual CUE2021, Sept 4–8, 2021. This paper is a substantial extension of the short version of conference paper. This work is supported by the National Natural Science Foundation of China (NSFC) (52007123) Corresponding authors: Xinwei Shen (sxw.tbsi@sz. tsinghua.edu.cn), Hongbin Sun (shb@tsinghua.edu.cn).

| | |
|---|---|
| $P_{s,t,i}^{FC}$ | Gas that the $i^{th}$ FC consumes at time $t$ in scenario $s$, unit: kW |
| $P_{s,t,i,e}^{FC}$ | Electricity output of $i^{th}$ FC at time $t$ in scenario $s$, unit: kW |
| $P_{s,t,i,h}^{FC}$ | Heating output of $i^{th}$ FC at time $t$ in scenario $s$, unit: kW |
| $P_{s,t,ch}^{B/TESS}$ | Charge power of ESS at time $t$ in s, unit: kW |
| $P_{s,t,dis}^{B/TESS}$ | Discharge power of ESS at time $t$ in s, unit: kW |
| $P_{s,t,j,ch}^{EV}$ | Charge power of the $j^{th}$ EV at time $t$ in s, unit: kW |
| $P_{s,t,j,dis}^{EV}$ | Discharge power of the $j^{th}$ EV at $t$ in s, unit: kW |
| $P_{s,t}^{PV}$ | Output of the PV at time $t$ in scenario s, unit: kW |
| $E_{s,t}^{B/TESS}$ | Energy stored in ESS at time $t$ in scenario s, unit: kWh |
| $E_{s,t,j}^{EV}$ | Energy stored in the $j^{th}$ EV at time $t$ in scenario s, unit: kWh |
| $\tilde{E}_{dep,s,j}^{EV}$ | Energy of the $j^{th}$ EV when it leaves in scenario s, unit: kWh |
| $\xi$ | Random vector |
| $Y_{s,t}^{ESS/EV}$ | Binary variable which indicates the charge state of ESS/EV |
| $Z_s$ | Binary variable which indicates the $s^{th}$ scenario is standard or not |

## I. INTRODUCTION

Climate change caused by environmental problems is a significant issue of common concern to all mankind. More than 70 countries have committed to working toward net-zero emissions by 2050 and to enhance their international climate pledges under the Paris Agreement to ensure that global warming is controlled below 2 degrees Celsius. Among them, China proposes to reach the peak of carbon dioxide emissions by 2030 and achieve carbon neutrality by 2060. In this context, renewable energy such as photovoltaic (PV) and wind energy has developed rapidly. Besides, the rapid development of equipment like fuel cell (FC), which could be fueled by clean energy, and able to supply both electricity and heat, will further accelerate the process of achieving zero carbon emission. Moreover, the penetration rate of electric vehicles (EV) is increasing. With the development of vehicle to grid (V2G) related technologies, EVs have great application potential in both reducing carbon emissions and increasing the flexibility of the grid. However, these may also bring more problems and challenges to the energy system. To cope with such problems and make better use of the strong coupling and synergy between different energy systems, the concept of Energy Internet has become another important topic in the energy industry after the Smart Grid [1]. As the physical carrier of the Energy Internet, Integrated Energy System (IES) breaks the boundary between different energy systems and promotes energy efficiency to the greatest extent. We believe that building-level IES which is represented by industrial parks and office/hospital/hotel buildings, will become the main energy carrier and play an important role in carbon neutrality in the future. Therefore, the planning problem of the building-level IES, which considers electricity-heat-V2G coupling, has become the research topic of this study. The planning problem of traditional energy systems like power system has been studied thoroughly, but there are still some research gaps that lie in the planning of IES, which will be analyzed from relevant studies as follows.

First of all, few studies have taken carbon emissions into account in the planning of IES, while we think this is one of the most important functions that building-level IES will undertake in the future. In addition, the immaturity of carbon price related mechanism also leads to different standards and methods. Multi-objective optimization which takes minimum cost and carbon emission as objective is commonly used in these studies [2, 3], but it is difficult for them to be optimal simultaneously. Wang et al. [4] introduce the external electricity ratio as the third objective which makes the objective more comprehensive but also makes the decision more difficult. Single objective planning takes the minimum total cost as the objective function is adopted too, Geng et al. [5] proposes a chance-constrained planning method for microgrid, although carbon emission is not directly considered, only renewable energy is used to provide electricity in a microgrid, so as to reduce carbon emission. Ge et al. [6] adopts a three-stage carbon trading price to convert carbon emission into operational cost and get the most economical planning scheme for IES. Bartolini et al. [7] reflects the change of carbon price on the change of energy price and analyzed the impact of carbon price change on IES planning results. Therefore, even for the same objective, the methods of considering carbon emissions vary greatly.

Secondly, the model formulation of building-level IES still needs to be studied further due to its higher complexity in terms of coupling different energy systems. This study adopts the Energy Hub (EH) model proposed by [8], which is widely used, and considers as many elements as possible to establish a comprehensive model. For IES, one of the most important equipment is multi-energy conversion equipment, in which different forms of energy are transformed and coupled with each other. Given the large capacity of CHP unit, which is not suitable for building-level IES, the FC which in smaller capacity and cleaner is used as the multi-energy equipment in this study. FC power systems offer a unique combination of high efficiency, wide size range, modularity, and compatibility with cogeneration [9]. Although system cost and durability remain as the major challenge for FC, related technologies develop fast thus FCs are wildly considered in IES planning studies [10, 11] and verified the effectiveness of FC as an energy conversion equipment in IES. In addition, in order to further strengthen the capacity of new energy consumption and solve the problem of time mismatch between renewable energy output and load demand, energy storage system (ESS) related technology has developed rapidly and also be widely considered in the planning problems of IES nowadays [12, 13]. So multi-energy storage system will also play an important role in this study. Moreover, with the rise of the number of EVs, their charging demand has constituted an important part of the load of power grid. Based on this, the applications of V2G technology have attracted extensive attention. Wang et al. [14], [15] apply the V2G technology in the planning problem of distribution network, the former fully analyze the stochasticity of EV traveling/charging behavior and its impact on the planning result of distribution network while the latter focus on the optimal decision scheme for site and capacity of EV charging stations. Waqar et al. [16] take the microgrid as research objection and analyze the effect of V2G on improving the reliability of energy supply and economy of the microgrid. Aluisio et al. [17] also consider a V2G-equipped microgrid and are mainly concerned with the optimal day-ahead operation of the microgrid. Pang et al. [18] proposes the conception of vehicle to building (V2b) based on V2G technology and analyze the peak shaving function of EVs on the building load. Wei et al. [19] consider V2G in building-level IES and deal with the uncertainty it brings. The development of V2G technology further explore the value of EVs and will be considered in this study. The above researches only consider the impact of one or a few factors on IES planning but ignore the integration nature of IES.

Besides, the IES has greater uncertainties than traditional energy systems which should be addressed properly by appropriate methods. Corresponding literature review on the stochastic planning research of energy system has been done on distribution network (DN), active distribution network (AND), microgrid, power system (PS) and IES, which can be found in **Table 1**. The existing studies do not fully consider the uncertainties of IES, and there is little consideration of the correlation between a variety of uncertainties.

To address the gaps in aforementioned researches, this study proposes a novel framework for planning problem of building-level IES, the main contributions are as follow:

(i) Formulate a chance-constrained two-stage (CCTS) stochastic planning model whose objective is to minimize the total cost including carbon tax based on a comprehensive IES model considering multiple loads, V2G interaction, FCs, renewable energy, ESS, etc.

(ii) The correlation of different uncertainties in building-level IES is considered by applying the heuristic moment matching scenario generation (HMMSG) method and the sample average approximation (SAA) method is combined with it to solve the CCTS model for the first time.

## II. FRAMEWORK OF PROPOSED PLANNING METHOD

The framework of the proposed planning method for building-level IES is shown as **Fig. 1**, it also reflects how the rest of paper will be organized. The planning model consists of three parts. First of all, the EH model based on equipment parameter is used to model our building-level IES. Then is the objective function, which minimizes the total cost including cost on investment, operation, carbon emission and the compensation for EV users who do not get enough electricity after participating in V2G bi-direction interaction. Additionally, there are constraints for power supply and equipment. The chance constraint is used to address the uncertainties. In this model we utilize chance constraints for the charge/discharge process of EVs to consider the uncertainties of EV users' behavior and the flexibility of EVs. The decision variables of this model are divided into the investment decision variables in the first stage and the operational decision variables in the second stage. This CCTS planning model is recast by SAA method and then solved with different scenarios generated from the history data by HMMSG method.

## III. MODEL FORMULATION AND SOLVING ALGORITHM

The EH model mentioned above is used to model the building-level IES in this work as mentioned before. It is a two-port model used to describe the relationship between energy input and output of IES. Since it was proposed in 2007 by the ETH Zurich in the project "vision of future energy network" [8], it has been widely used in the modeling of IES and its effectiveness has been verified. One of the biggest advantages of EH model is its simplicity: different forms of energy are put into the energy hub and then converted into different forms of energy output to the load side. The mathematical model can be shown as:

$$\mathbf{O} = \mathbf{CI} \qquad (1)$$

We use $\mathbf{O}$ and $\mathbf{I}$ to represent the input energy and output energy of IES respectively, $\mathbf{C}$ is the coupling matrix which reflects the coupling relationship, such as the efficiency of energy conversion equipment and the specific parameters of energy transmission network between input and output. With this coupling matrix, we can only focus on the input and output energy of IES. Another advantage of the EH model is that it can be adjusted easily according to specific requirement. When the ESS is considered in IES, the mathematical model can be modified as follow[35]:

$$\mathbf{O} = \mathbf{CI} - \mathbf{S} \quad (2)$$

**Table 1**
Relevant research review

| Ref. | Research object | Modelling techniques | Problem category | Uncertain parameters under consideration | | | |
|------|-----------------|----------------------|------------------|------|------------------|-----|--------|
|      |                 |                      |                  | Load | Renewable energy | EV  | Others |
| [5]  | Microgrid       | Chance constraint    | Planning         | √    | √                |     |        |
| [6]  | IES             | Affined model        | Planning         |      | √                |     |        |
| [20] | IES             | Scenario generation  | Planning         | √    | √                |     |        |
| [21] | IES             | Robust optimization  | Planning         | √    |                  |     |        |
| [22] | IES             | Scenario tree        | Planning         | √    |                  |     | √      |
| [23] | IES             | Scenario reduction   | Planning and risk assessment | √ | √ |     |        |
| [24] | DN              | Probabilistic model  | Planning         | √    | √                |     |        |
| [25] | DN              | Chance constraint    | Planning         | √    | √                |     |        |
| [26] | DN              | Scenario tree        | Planning         | √    | √                |     | √      |
| [27] | DN              | Scenario generation  | Planning         | √    | √                |     |        |
| [28] | DN              | Chance constraint    | Planning         |      | √                | √   |        |
| [29] | ADN             | Typical scenario     | Planning         | √    |                  |     |        |
| [30] | ADN             | Chance constraint    | Planning         | √    | √                |     |        |
| [31] | ADN             | Scenario generation  | Planning         | √    | √                |     |        |
| [32] | PS              | Chance constraint    | Optimal power flow | √  |                  |     |        |
| [33] | PS              | Chance constraint    | Unit Commitment  |      | √                |     |        |
| [34] | PS              | Chance constraint    | Transmission line expansion planning | | √ | | |

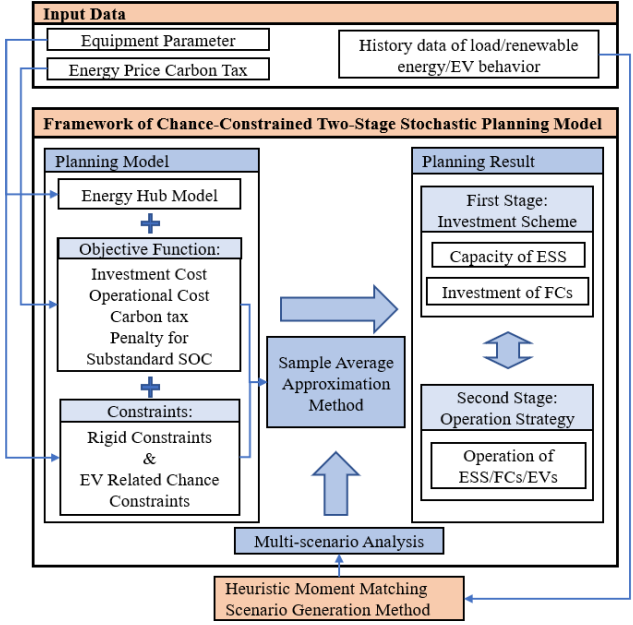

Fig. 1. The framework of proposed planning method.

$\mathbf{S}$ represents the total output (charge/discharge power) of ESS at every time slot, when EV and some other elements is integrated into the EH, we can modify the model accordingly. For instance, EV can be regarded as a battery ESS (BESS) and considered in $\mathbf{S}$ when it is static.

*A. Proposed Planning Model of Building-level IES*

Based on the EH model, we formulate a CCTS stochastic planning model for building-level IES considering ESS, FCs, PV and EVs in this part. Relevant uncertainties are considered.

1) Objective function

For most IES planning problems, the main goal is to minimize the total cost or maximize the total income of IES. As mentioned above, this study also takes reducing carbon emission as one of the goals to achieve carbon neutrality. We combine these two goals into a single objective function, while the carbon emission is converted into "carbon tax" and counted as part of the total cost. The objective function of the planning model is shown as follows:

$$\min f^{inv}(\mathbf{X}) + \mathrm{E}\left[f^{ope}(\mathbf{P},\xi) + f^{carbon}(\mathbf{P},\xi) + f^{SOC}(\mathbf{P},\xi)\right] \quad (3)$$

in which

$$f^{inv} = X^{ESS} \cdot IC^{ESS} + \sum_{i \in \Omega} X_i^{FC} \cdot IC_i^{FC} \quad (4)$$

$$f^{ope} = m \sum_{s=1}^{N} \sum_{t=1}^{T} (\sum_{i \in \Omega} P_{s,t,i}^{FC} \cdot r_i^g + P_{s,t}^{Grid} \cdot r_t^e) \quad (5)$$

$$f^{carbon} = m \cdot C^{tax} \sum_{s=1}^{N} \sum_{t=1}^{T} (\sum_{i \in \Omega} \varepsilon_i^g \cdot P_{s,t,i}^{FC} + \varepsilon_t^e \cdot P_{s,t}^{Grid}) \quad (6)$$

$$f^{SOC} = m \sum_{s=1}^{N} \sum_{j=1}^{NEV} p(0.9 - \tilde{E}_{dep,s,j}^{EV} / E_{\max}^{EV}) \quad (7)$$

$$\Omega = \{\text{SOFC}, \text{PEM}_{gas}, \text{PEM}_{H_2}\}$$

where the investment cost, operational cost, carbon emission cost and the penalty for the insufficient state of charge (SOC) of EVs are denoted by $f^{inv}$, $f^{ope}$, $f^{carbon}$ and $f^{SOC}$ respectively. $\mathbf{X}$ is the vector of the first-stage investment decision variables which decide the construction scheme of equipment, i.e., $\mathbf{X}=[X^{ESS}, X_i^{FC}, ...]$. $\mathbf{P}$ consists of second-stage operation decisions which represent the operation scheme of invested equipment, i.e., $\mathbf{P}=[[P_{s,t,i}^{FC},...]^T,[P_{s,t}^{Grid},...]^T,...]$. $\xi$ denotes the uncertain parameters' vector so the second part of the objective function related to it should be written in the expectation form as in Eq. (3). Eq. (4) shows that the investment cost includes the investment of ESS and all kinds of FCs, in which the capacity of ESS is denoted by $X^{ESS}$ and the number of FCs is denoted by $X_i^{FC}$, with their unit cost denoted by $IC^{ESS}$ and $IC_i^{FC}$. Three kinds of FCs are taken as options which are shown in set $\Omega=\{\text{SOFC}, \text{PEM}_{gas}, \text{PEM}_{H_2}\}$, including polymer electrolyte membrane fuel cell (PEMFC) consuming hydrogen denoted by $\text{PEM}_{H_2}$, solid oxidize FC (SOFC) and $\text{PEM}_{gas}$ which consume relatively cheaper fuel, natural gas. Operational cost and carbon cost are shown as Eq. (5) and Eq. (6) respectively. They are both related to the energy that consumed by different equipment in different scenarios, to be more specific, the gas that FCs consumed and the electricity imported from the grid. $P_{s,t,i}^{FC}$ means the gas that consumed by the $i^{th}$ kind of FC at time $t$ in scenario $s$, $P_{s,t}^{Grid}$ means the electricity that bought from the grid at time $t$ in scenario $s$. The constant m converts all the operation costs in the planning period into the planning year through the discount rate of cash $\gamma$, it can be expressed as follow:

$$m = \sum_{y=1}^{PP} \frac{365}{N \times (1+\gamma)^{t-1}} \quad (8)$$

where $N$ represents the number of scenarios. Eq. (7) is a penalty term for insufficient SOC levels of EVs. $NEV$ denotes the number of EVs. If the EV participating in V2G interaction does not reach the SOC level its owner expected when it leaves the building, the EV owner should be compensated. $p$ denotes the penalty for EVs those leave without expected SOC level, set as 0.9, and $\tilde{E}_{dep,s,j}^{EV}$ is the energy of the $j^{th}$ EV when it leaves in scenario $s$.

It should be noted that our proposed model converts emissions into cost and adds them to the objective function. It has two parts, the first part is related to FCs that consume natural gas, which means SOFC and $\text{PEM}_{gas}$. The second part is related to the electricity that bought from the grid. To the best of our knowledge, the electricity provided by the grid comes from different generations such as thermal power plants, hydropower plants, nuclear power plants, renewable energy output and so on in different proportions. In different periods within a day, the proportion of different generations' power is different, which means consuming the same amount of electricity at different time of a day equals to different carbon emission intensity. Therefore in Eq. (6), the carbon emission coefficient of electricity from the grid $\varepsilon_t^e$ has different value at different time *t*. In fact, currently the power grid will not charge users for additional fees on these carbon emissions, and there is also no relevant policy to subsidize users for purchasing electricity with lower emission. However, in order to further explore the role of building-level IES in reducing carbon emissions and accelerating the realization of carbon neutrality, an additional carbon tax is set on the electricity purchased from the grid in this model. This is also a rational assumption on future energy policy considering the key role of building-level IES in emission reduction.

2) Energy balance constraints

The first group of constraints is the supply-demand balance constraint which can be obtained from formula (2):

$$\mathbf{L} \leq \mathbf{C}^{Equi} \times \mathbf{P}^{Equi} - \mathbf{\eta}^{ESS/EV} \times \mathbf{P}^{ESS/EV} \quad (9)$$

$$L_{s,t}^e = C_{l-e}^{PV} \cdot P_{s,t}^{PV} + \sum_{i \in \Omega} C_{g-e,i}^{FC} \cdot P_{s,t,i}^{FC} + C_{e-e}^{Grid} \cdot P_{s,t}^{Grid} - P_{s,t,ch}^{BESS} / \eta_{ch}^{BESS} + \eta_{dis}^{BESS} \cdot P_{s,t,dis}^{BESS} - \sum_{j=1}^{NEV}(P_{s,t,j,ch}^{EV}/\eta_{ch}^{EV} - \eta_{dis}^{EV} P_{s,t,j,dis}^{EV}) \quad (10)$$

$$L_{s,t}^h \leq \sum_{i \in \Omega} C_{g-h}^{FC} \cdot P_{s,t,i}^{FC} - \sum_{i \in \Omega} P_{s,t,ch}^{TESS}/\eta_{ch}^{TESS} + P_{s,t,dis}^{TESS} \cdot \eta_{dis}^{TESS} \quad (11)$$

$\mathbf{L} = [\mathbf{L}^e = [L_{s,t}^e, ...]^T, \mathbf{L}^h = [L_{s,t}^h, ...]^T]^T$ denotes the load of IES. $\mathbf{P}^{Equi}$ denotes the output of different equipment such as PV and FCs, i.e. $\mathbf{P}^{Equi} = [[P_{s,t}^{PV}, ...]^T, [P_{s,t,i}^{FC}, ...]^T, [P_{s,t}^{Grid}, ...]^T]^T$. $\mathbf{C}^{Equi}$ represents the coupling matrix of different equipment and the subscripts indicate different kinds of energy, i.e. $\mathbf{C}^{Equi} = [[C_{l-e}^{PV}, C_{l-h}^{PV}]^T, [C_{g-e}^{FC_i}, C_{g-h}^{FC_i}]^T, [C_{e-e}^{Grid}, C_{e-h}^{Grid}]^T]$. As for electricity, the supply must be equal to the demand at any time under all scenarios (without considering power losses), but for heat system, the supply can be equal or larger than the demand as Eq. (10) and Eq. (11) indicate. BESS and TESS represent battery ESS and thermal ESS respectively. $\mathbf{\eta}^{ESS/EV}$ denotes the matrix of charge or discharge efficiency of ESS/EVs, i.e. $\mathbf{\eta}^{ESS/EV} = [[\frac{1}{\eta_{ch}^{BESS}}, 0]^T, [\eta_{dis}^{BESS}, 0]^T, [0, \frac{1}{\eta_{ch}^{TESS}}]^T, [0, \eta_{dis}^{TESS}, 0]^T, [\frac{1}{\eta_{ch}^{EV}}, 0]^T, [\eta_{dis}^{EV}, 0]^T]$. $\mathbf{P}^{ESS/EV}$ denotes the matrix of charge/discharge power of BESS, TESS and EVs. The subscript *ch* and *dis* demote charge and discharge respectively.

3) Output limits of FC and substations

The second group of constraints are the output limitations of different equipment. Eq. (12)—(14) show the power limits for PVs, FCs and substation. The power outputs of FCs are related to their investment decision variables, namely $X_i^{FC}$:

$$0 \leq P_{s,t}^{PV} \leq P_{max}^{PV} \quad (12)$$

$$0 \leq P_{i,s,t}^{FC} \leq X_i^{FC} \cdot P_{i,max}^{FC}, i \in \Omega \quad (13)$$

$$0 \leq P_{s,t}^{Grid} \leq P_{max}^{Grid} \quad (14)$$

4) ESS-related constraints

The next important set of constraints is for BESS:

$$SOC_{min}^{BESS} \leq E_{s,t}^{BESS} / X^{ESS} \leq SOC_{max}^{BESS} \quad (15)$$

$$E_{s,t}^{BESS} = E_{s,t-1}^{BESS} + P_{s,t-1,ch}^{BESS} - P_{s,t-1,dis}^{BESS} \quad (16)$$

$$P_{s,t,ch/dis}^{BESS} \leq \alpha^{BESS} \cdot X^{ESS} \quad (17)$$

$$E_{s,1}^{BESS} = E_{s,T}^{BESS} \quad (18)$$

$$\sum_{y=1}^{PP} 365 \times \sum_{t=1}^{T} P_{s,t,ch}^{BESS} \leq T_{ch}^{BESS} \cdot X^{ESS} \quad (19)$$

$$P_{s,t,ch}^{BESS} \leq M \cdot Y_{s,t}^{BESS} \quad (20)$$

$$P_{s,t,dis}^{BESS} \leq M \cdot (1 - Y_{s,t}^{BESS}) \quad (21)$$

where $E_{s,t}^{BESS}$ denotes the amount of energy stored in the ESS at time *t* in scenario *s*, $X^{ESS}$ is investment decision variable which indicates the capacity of ESS to be invested. Eq. (15) is a constraint for SOC of battery, the lower and upper limits are usually set as 0.1 and 1 to maximize its life. Eq. (16) reflects the variation of energy stored in BESS, since the efficiency have been considered in the EH model, Eq. (10), it is not necessary to consider it here. In Eq. (17) $\alpha^{BESS}$ denotes the coefficient converting the BESS capacity to its charge/discharge rate. For instance, $\alpha^{BESS} = 0.25$ if we consider to invest BESS with a maximum capacity of 4MW and its maximum charging power is 1MW. Eq. (18) shows that the energy stored in the ESS has to be equal at the beginning and end of each day. Constraint (19) limits the total charge/discharge energy of BESS in the whole planning period. $T_{ch}^{BESS}$ means the maximum full charge/discharge cycles of BESS, which is set as 5000 in our model. Constraint (20) and (21) mean that the BESS cannot charge and discharge at the same time. *M* is a big constant and $Y_{s,t}^{BESS}$ is a binary variable which denotes the charge state of BESS at time *t* in scenario *s*. The constraints for thermal ESS (TESS) are similar so we are not going to repeat them here.

5) EV-related constraints

$$SOC_{min}^{EV} \leq E_{s,t}^{EV} / E_{max}^{EV} \leq SOC_{max}^{EV} \quad (22)$$

$$E_{s,t,j}^{EV} = E_{s,t-1,j}^{EV} + P_{s,t,j,ch}^{EV} - P_{s,t,j,dis}^{EV}, t \in \Psi_{parking} \quad (23)$$

$$P_{s,t,j,ch}^{EV} \leq P_{charger}^{EV} \quad (24)$$

$$P_{s,t,j,dis}^{EV} \leq \alpha^{EV} \cdot E_{max}^{EV} \quad (25)$$

$$P_{s,t,j,ch}^{EV} \leq M \cdot Y_{s,t,j}^{EV} \quad (26)$$

$$P_{s,t,j,dis}^{EV} \leq M \cdot (1 - Y_{s,t,j}^{EV}) \quad (27)$$

EVs' operation states are categorized as commuting hours and parking hours. When used for commuting, EVs cannot be charged or discharged, so we only considered the parking hours of EVs ($t \in \Psi_{parking}$). When EVs are parked in the building, they are regarded as BESS and limited by constraints (22)—(27) which are similar to the constraints for BESS. What makes EVs different from BESS is that they will be driven away when their users get off work.

To ensure the normal use of EVs, constraint requests that the SOC of EVs at their departure time should not be less than 0.9. However, it is hard to formulate this as a rigid constraint like what have been done to limit the output of other equipment due to the great uncertainty of EV users' behavior. Besides, in this building-level IES or any other energy system, load curtailment is considered as the last thing that we want to see. So, the supply-demand balance constraints should be satisfied in any scenarios as rigid constraints and we only formulate the departure SOC limitation for EVs in chance constraint form.

Reference [36] firstly introduces chance-constrained programming and two essential ways of writing chance-constrained model, namely single separated chance constraint and joint chance constraint, which can be indicated as follow formulas respectively.

$$p_i(x) := \Pr\{G_i(x,\xi) \leq 0\} \geq 1-\varsigma_i \quad i=1,2,...,n \quad (28)$$

$$p(x) := \Pr\{G_1(x,\xi) \leq 0,...,G_n(x,\xi) \leq 0\} \geq 1-\varsigma \quad (29)$$

The first constraint means that the probability that the inequality $G_i(x,\xi) \leq 0$ is satisfied is greater than $1-\varsigma_i$ and there will be $n$ constraints. The second constraint requires that the probability of all $n$ constraints being satisfied simultaneously is greater than $1-\varsigma$, which is much stricter. Then we can write our SOC chance constraint in joint form as follow:

$$\Pr\left\{0.9 - \tilde{E}_{dep}^{EV} / E_{max}^{EV} \leq 0\right\} \geq 1-\varsigma \quad (30)$$

It suggests that all EVs have to leave with specified SOC level with probability larger than $1-\varsigma$, which is much stricter and applied in the proposed model.

To conclude, the proposed CCTS stochastic planning model for building-level IES is as below:

$$\begin{aligned} &\min_{X,P,Y,\xi} (3) \\ &\text{s.t. } \{(9), (12)-(27), (30)\} \end{aligned} \quad (\text{P 1})$$

### B. Solving Algorithm for CCTS Stochastic Planning Model

Since many uncertainties are considered and their probability distribution is difficult to be described, the chance constraint needs to be transformed into a form that is easy to be solved. Referring to [33], the SAA method based on multi-scenario is an effective method to solve CCTS problems. A brief introduction of this method will be given in this part. The key of SAA is to generate a large enough number of scenarios, then use the average of samples to estimate the expectation value in the objective function and the satisfaction probability of chance constraints. Then the objective function can be reformulated as

$$\min f^{inv}(\mathbf{X}) + \frac{1}{N}\sum_{s=1}^{N}(f_s^{ope} + f_s^{carbon} + f_s^{SOC}) \quad (31)$$

In order to reformulate the chance constraint, an indicator function $\mathbf{1}_{(0,\infty)}(\cdot)$ is introduced as Eq. (32), and then constraint (30) is equivalent to Eq. (33) as follow:

$$\mathbf{1}_{(0,\infty)}(G(x)) = \begin{cases} 0, & G(x) \leq 0 \\ 1, & G(x) > 0 \end{cases} \quad (32)$$

$$N^{-1}\sum_{s=1}^{N}\mathbf{1}_{(0,\infty)}(0.9 - \tilde{E}_{s,dep}^{EV} / E_{max}^{EV}) \leq \varsigma \quad (33)$$

The indicator function can also be recast by big M method as we have mentioned before. A group of binary variables $Z$ is added into the model, then the joint chance constraint is reformulated to a group of deterministic constraints which are much easier to be solved as follows:

$$0.9 - \tilde{E}_{s,dep}^{EV} / E_{max}^{EV} \leq M \cdot Z_s \quad (34)$$

$$\sum_{s=1}^{N} Z_s \leq N \cdot \varsigma \quad (35)$$

As a result, the CCTS model is reformulated as a mixed-integer linear programming (MILP) model as follow:

$$\begin{aligned} &\min_{X,P,\xi,Y,Z} (31) \\ &\text{s.t. } \{(9), (12)-(27), (34), (35)\} \end{aligned} \quad (\text{P 2})$$

which can be solved directly by state-of-art commercial solver, e.g., CPLEX or GUROBI.

## IV. CASE STUDY

To verify the effectiveness of the proposed planning model, a real building in Shanghai is tested as the numerical example. For this building-level IES, energy inputs include gas, electricity and solar energy, the demands include electricity, heating and cooling energy. The equipment considered to be invested include ESS and FCs. As discussed earlier, three kinds of FCs that consume different fuels are candidate options. Solar panels and EVs are considered as existing equipment, which we mainly focus on their operation optimization without considering their investment but not investment decisions. All parameters of these equipment are shown in **Table 2** and **Table 3**. In this sector, an HMMGS method will be introduced first. Then we will make a specific analysis on the planning result and the application scenarios and significant roles of the ESS/EVs based on this real planning case.

### A. Case Conditions and Scenario Generation

Since SAA is a scenario-based method, the key of it is the effectiveness of scenario generation method. We consider uncertainties including different types of loads, output of PV and the random behaviors of EV users and their correlation. The original data includes the load of another similar building-level IES in the same area, the output of PV and EV user's behavior including the arriving/departure time of them and the initial SOC level at arriving time of 8760h. Based on them, the HMMSG method proposed by [37] is applied to generate scenarios. HMMSG does not require random parameters follow specific distribution, e.g., normal/Beta distribution etc. Instead, it generates scenarios by matching the first four moments and the correlation matrix between random variables with information from historical data sets. We generate 100 scenarios with HMMSG based on the original data sets. Energy price is shown as follows: Shanghai's time-of-use electricity price is adopted and the price of natural gas is ¥2.57/m$^3$. At present, the carbon emission tax in Shanghai is about ¥40/ton of carbon dioxide. According to the Paris Agreement, the carbon tax should reach the US $40—80 by 2020 and US $50—100 by 2030. Although the carbon tax is far from meeting the expectations, we set the carbon tax at different levels to observe the effect on the planning results. For chance constraints, $\varsigma$ is set as 0.05. The penalty for EV with substandard SOC is ¥1/kWh.

**Table 2**
Related parameters of FCs.

| FCs type | $IC_i^{FC}$ (¥$10^4$) | $C_{g-e}$ | $C_{g-h}$ | $P_{e,max}^{FC}$ (kW) | $P_{h,max}^{FC}$ (kW) |
|---|---|---|---|---|---|
| SOFC | 80 | 0.63 | 0.28 | 4.5 | 2 |
| $PEM_{gas}$ | 30 | 0.34 | 0.5 | 4.2 | 6.2 |
| $PEM_{H_2}$ | 295 | - | - | 60 | 60 |

**Table 3**
Related parameters of other equipment.

| | $IC^{ESS}$ (¥$10^4$/kWh) | $P_{ch,max}^{ESS/EV}$ | $P_{dis,max}^{ESS/EV}$ | Capacity |
|---|---|---|---|---|
| BESS | 0.15 | 0.25 $X^{ESS}$ | 0.25 $X^{ESS}$ | - |
| TESS | - | 0.5 $E_{max}^{TESS}$ | 0.5 $E_{max}^{TESS}$ | 150kWh |
| EV | - | 7kW | 0.25 $E_{max}^{EV}$ | 60kWh |
| PV | - | - | - | 1000kW |

*B. Analysis on Planning Results and Operation Strategy in Extreme Scenarios*

The planning result under different carbon taxes, from ¥40 RMB/ton to ¥1000 RMB/ton, are shown in **Table 4**. It can also be regarded as the first stage result of our CCTS model. It can be seen that although the gas-to-electric efficiency of SOFC in this study is relatively high, it will not be installed in any case because of its high investment cost and relatively low gas-to-heat efficiency. Moreover, the choice of PEM is greatly affected by the carbon tax since they consume different fuels. Due to the high investment of PEM that uses hydrogen, when the carbon tax is relatively low, the building-level IES prefers $PEM_{gas}$. When the carbon tax increases, the number of $PEM_{H_2}$ gradually increases while $PEM_{gas}$ decreases. The number of "substandard scenarios", in which EVs' SOC level is below their owners' expectations, is also significantly affected by the carbon tax. The number of substandard scenarios is increasing from 0 to 5 with the increment of carbon tax. This can be explained easily by the carbon tax levied on electricity bought from the grid. Although there is a penalty for substandard SOC of EVs, it will be ignored when we can benefit more from avoiding high carbon tax. It should be noted that even it can profit a lot from the rising carbon tax, the number of substandard scenarios should not exceed 5 due to the chance constraints on EVs ($\varsigma$ =0.05 in 100 scenarios). The capacity of invested BESS changes slightly with the variation of the carbon tax, about 1.5 MWh BESS will be invested for energy arbitrage.

**Table 4**
The Planning result under different carbon tax level

| Carbon tax/¥ | SOFC/sets | $PEM_{gas}$/sets | $PEM_{H_2}$/sets | BESS /kWh | Substandard scenarios |
|---|---|---|---|---|---|
| 40 | 0 | 27 | 0 | 1490 | 0 |
| 100 | 0 | 17 | 1 | 1490 | 0 |
| 400 | 0 | 7 | 2 | 1400 | 1 |
| 700 | 0 | 7 | 2 | 1520 | 4 |
| 1000 | 0 | 0 | 3 | 1330 | 5 |

**Table 5** shows the cost details under different carbon tax. For operational cost, the increase of carbon tax leads to more gas purchase costs. This is due to the increased purchase of hydrogen which is more expensive but cleaner than the natural gas. Larger amount of hydrogen is used to support the operation of $PEM_{H_2}$ and reduce carbon emissions. This also decrease the cost on electricity that bought from the grid. The cost on carbon emission will increase for sure with the raise of tax. As for the penalty for substandard SOC level, the reason for its increase is the same as the reason for the increase of substandard scenarios mentioned above. Although the price of ¥1000/ton is rather high and cannot be reached in a short time, with the increasingly strict requirements for environmental protection and the development of carbon trading mechanism, this price will be reasonable in the future.

Then we will focus on the optimal operational results of the stochastic planning model, which can also be regarded as the second stage result of our CCTS model, based on the planning scheme under carbon tax of ¥400/ton. In order to show the basic operation of the equipment under this planning scheme, two extreme scenarios with maximum electricity load and maximum heating load respectively are chosen from 100 scenarios, **Fig. 2** depicts the operation strategy of each equipment in these 2 extreme scenarios. It can be seen that the electricity load is supplied by the output of multiple equipment including the grid, BESS, different kinds of FCs, EVs and PV. The area that beyond the load curve is electricity used for charging BESS and EVs. And the heating load is supplied by the output of TESS and two kinds of PEM.

**Table 5**
Cost of stochastic planning result under different carbon tax (unit: ¥ $10^4$RMB).

| Carbon tax (¥/ton) | Investment Cost | | Operational Cost | | Carbon Emission Cost | | SOC penalty | Total Cost |
|---|---|---|---|---|---|---|---|---|
| | On FCs | On BESS | Gas bought from outside | Electricity bought from the gird | From electricity | From gas | | |
| 40 | 810 | 224 | 199 | 1112 | 57 | 7 | 0 | 2409 |
| 100 | 805 | 224 | 227 | 1094 | 140 | 12 | 0 | 2502 |
| 400 | 800 | 211 | 573 | 863 | 460 | 14 | 0.4 | 2921.4 |
| 700 | 800 | 228 | 589 | 838 | 794 | 16 | 5.0 | 3270.0 |
| 1000 | 885 | 199 | 793 | 711 | 986 | 0 | 6.2 | 3580.2 |

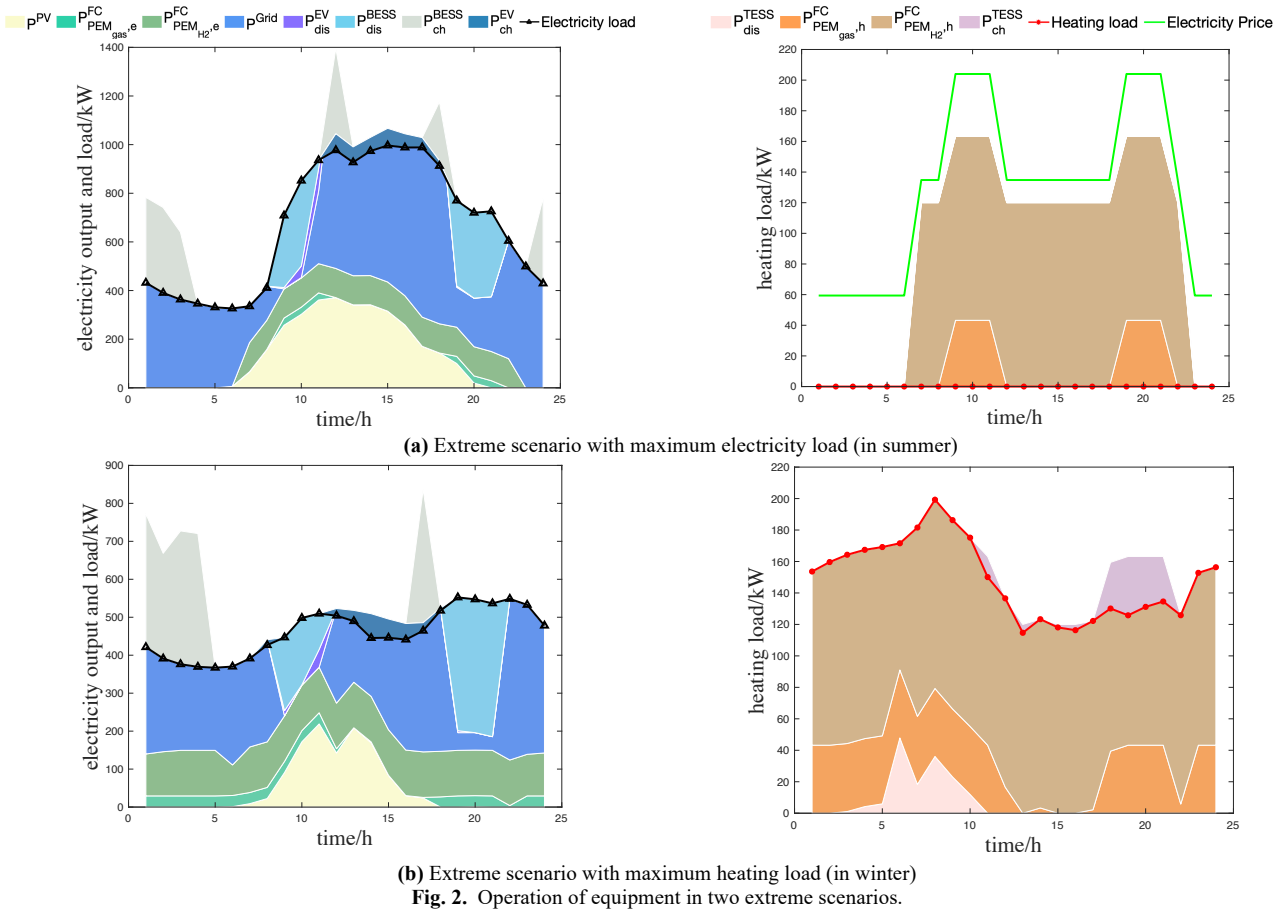

**(a)** Extreme scenario with maximum electricity load (in summer)

**(b)** Extreme scenario with maximum heating load (in winter)
**Fig. 2.** Operation of equipment in two extreme scenarios.

The scenario that with maximum electricity load can actually be regarded as an extreme summer day, there is no heating load and the output of photovoltaic is supplied from around 5 am to 8 pm. It can be seen that although there is no heating load, the FCs will still work in specific period, this can be explained by the high cogeneration efficiency of FCs. In this period, even if the generated heat is wasted, the per-unit cost of generating electricity is still lower than price of electricity from the grid. Especially in peak price period, two kinds of FCs cogenerate together to reduce total costs.

The extreme scenario with maximum heating load can be considered as an extreme winter day for the same reason. The output of photovoltaic is relatively small and supplied only from around 6 am to 18 pm in this day, which also proved the effectiveness of HMMSG on considering correlations between different data sets (multi-energy load and PV power outputs). In this scenario, the FCs work almost all the time because of the consistent demand for heating energy.

From both scenarios we can see that BESS reshape the load curve of electricity to pursue a maximum profit. As for heating load, sometimes the output of the PEM exceeds the heating load, TESS stores this part of heat and releases it in other periods, also in order to obtain the maximum benefit. We will further analyze the function of ESS in next two parts.

### C. Analysis on BESS and TESS in Typical Scenarios

This part also takes the planning results obtained under the condition that carbon tax in ¥400/ton as example. The function of BESS can be further illustrated by **Fig. 3** which shows the details of charging/discharging process of BESS in 3 random scenarios from 100 scenarios. It can be seen from the electricity price curve that the time-of-use electricity price in a day is divided into three stages: peak, average and valley price based on time of the day. The electricity prices of these 3 stages are 1.02, 0.674 and 0.297 ¥/kWh respectively. For BESS, purchasing electricity at valley hours and using stored energy to meet the load at peak hours will gain great benefits, namely, energy arbitrary. Therefore, the gray bars representing the charging power of BESS are mainly concentrated in the valley price period, from 22 pm to 5 am the next day. And all energy that discharge from the BESS to supply the building load, which represented by light blue bars, are concentrated in peak price hours to maximum profits. Although there is also potential profit from purchasing electricity at average hours to supply load in peak hours, the price gap is not big enough yet considering lifespan of BESS is limited and the charging/discharging losses, thus less charging happened at average hours.

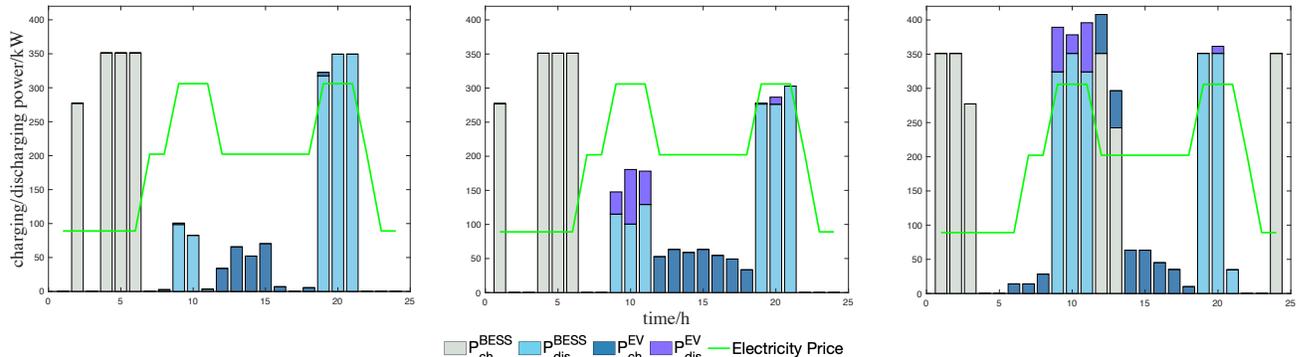

**Fig. 3.** Charging and discharging process of BESS/EVs and price curve of electricity.

As for heating storage, the existing TESS of 150KWh is used for charging and discharging, **Fig. 4** presents this process. The charging and discharging period of TESS are related to electricity price and also dependent on the heating load curve. Three scenarios with small heating load, medium heating load and maximum heating load are selected to observe the operation of TESS. For charging process, the PEM is likely to produce electricity and heat simultaneously during peak and average price period especially the second peak period, and the wasted heat will be stored by TESS. The reason that there is no charging process during the first peak can be explained by the load curve: this peak of electricity price is also the peak of heating load, so there is no excess energy for charging. As for discharging process, it prefers to release stored heating energy in valley periods, but it is also significantly affected by the heating load curve. When the heating load is rather large as (c) shows, discharge process even occurs in peak period due to the huge heating load demand.

In general, the operation of BESS and TESS are related to time-of-use electricity price. They both utilize the price gap between different periods to obtain the maximum profit. BESS prefers to charge at the valley price and discharge at the peak price, while TESS is on the contrary. In some cases, they will also charge and discharge energy at the average electricity price to make a profit. It should be noted that the operation of TESS is also closely related to the heating load. In addition, when the capacity of TESS is adjusted, the construction and operation of PEM will change accordingly, because TESS can optimize the operation of PEM and improve its benefits, so as to maximize the profits of whole IES.

### D. Analysis on EVs Based in Standard/Substandard Scenarios

In all 100 scenarios, there is only 1 scenario that have EVs that fail to meet the SOC level requirement at departure time under the carbon price of ¥400/ton. In order to observe the variation of SOC level of 10 EVs in this scenario, **Fig. 5** shows it and randomly selects 1 standard scenario for display. Firstly, the charging and discharging process of EVs will be analyzed. In **Fig. 3**, we can see that the charging power of EVs represented by dark blue bars concentrate in the period of average electricity price, while the discharging of EVs represented by purple bars appear in the first peak of electricity price. In **Fig. 5**, in different scenarios, each EV arrives at the building at different hours with different SOC levels. **Fig. 3** shows that most EVs discharge in the first peak price period because most of them arrive before the peak price begins. Then they are charged during the average price period to meet the SOC requirements at departure time, and it is easy to see that only a few discharged in the second peak electricity price stage. This is because the second peak price period is also when most of EVs leave the building.

The arrival time, departure time and SOC level of EVs in different scenarios are different, but they all reach the SOC level of 0.9 when leaving in the standard scenario as **Fig. 5** (a) shows. But the substandard scenario shown by **Fig. 5** (b) is opposite. The first reason that may lead to the violation is to leave too early, it can be seen from **Fig. 5** (b) that three EVs leave early and there is no enough time for charging. In the same figure, there are also some EVs violate the constraint which leads to another reason, which is to leave later than the end of the second peak price period. It can be seen that these EVs are charged to a high SOC level before leaving, and since the departure time is just after the peak price period, the profit obtained by using it for energy arbitrage may exceed the penalty, so the EVs choose to violate the constraint. This is a special scenario with EVs that either leave early or late. With the increase of carbon price, EVs can benefit more by V2G interactions, and there will be more substandard scenarios. On the one hand, we prevent the occurrence of too many substandard scenarios by introducing the chance constraint. On

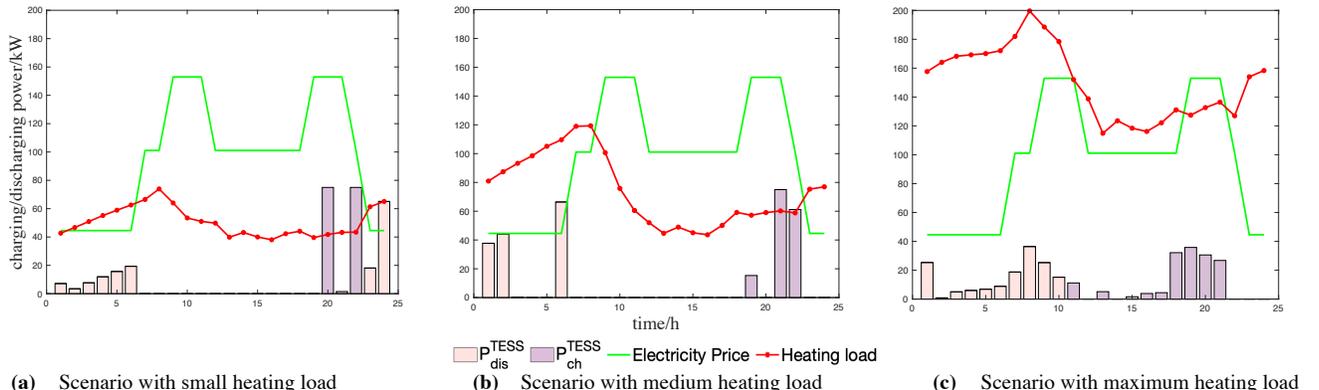

(a) Scenario with small heating load     (b) Scenario with medium heating load     (c) Scenario with maximum heating load

**Fig. 4.** Charging and discharging process of TESS, heating load curve and price curve of electricity.

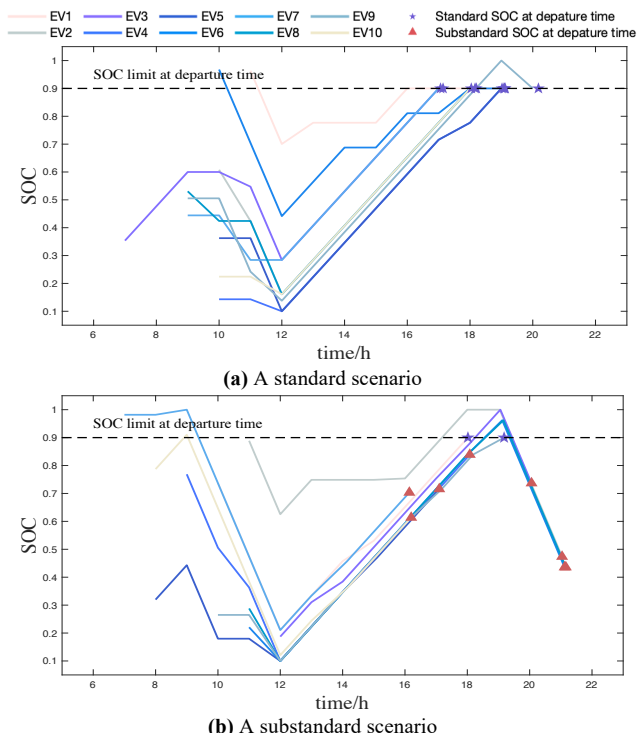

**Fig. 5.** SOC level variation of EVs in two scenarios.

the other hand, necessary measures and optimal V2G control method need to be taken to ensure the EVs that either leave early or late have sufficient SOC to provide safety and convenience for EVs' owners.

## V. CONCLUSION

This paper focuses on the planning problem of building-level IES and proposes a CCTS planning model to minimize the total cost and carbon emission of the IES. In this model, we consider the uncertainty from photovoltaic output, multi-energy load fluctuation and EV behavior, obtain the optimal investment scheme of ESS and FCs, as well as corresponding operation strategy. In addition, we focused on utilizing the flexibility of EVs, that is, V2G bi-direction interaction, and formulate the satisfaction of EV owners with chance constraints.

With the numerical case, the effectiveness of proposed model has been proved and we could conclude that the FC fueled by hydrogen which is cleaner will play a more important role in the future when carbon emission reduction is required. Besides, multiple ESSs are also proved to be key facilities in the operation of building-level IES. BESS can bring great benefits through energy arbitrary and reducing carbon emission with time-of-use electricity price and carbon emission coefficient. As for TESS, it can solve the time mismatch between supply and demand and greatly improve the operation efficiency of IES. With V2G capability, EVs also provide similar value without affecting EV owners' satisfaction.

However, there are still some limitations of this study such as the negligence for the modeling of network topology and some subjective assumptions about carbon trading. In addition, the different preferences of EV users are not distinguished in this study. The planning model should be further improved in the future to make it more accurate and generally applicable.

With the improvement of carbon trading mechanism, future research should also focus on its impact on planning for low-carbon building-level IES which considers EVs.